\documentclass[a4paper,noarxiv,twocolumn,accepted=2022-11-09]{quantumarticle}
\usepackage{amssymb}
\usepackage{graphicx}
\usepackage{amsmath}
\usepackage[colorlinks=true,linkcolor=blue,citecolor=blue,urlcolor=blue]{hyperref}

\begin{document}

\title{Quantized Nonlinear Transport with Ultracold Atoms}

\author{Fan Yang}
\affiliation{Institute for Advanced Study, Tsinghua University, Beijing 100084, China}
\author{Hui Zhai}
\email{hzhai@tsinghua.edu.cn}
\affiliation{Institute for Advanced Study, Tsinghua University, Beijing 100084, China}
\date{\today}

\begin{abstract}
In this work we propose a protocol to measure the quantized nonlinear transport using two-dimensional ultracold atomic Fermi gases in a harmonic trap. This scheme requires successively applying two optical pulses in the left and lower half-planes and then measuring the number of extra atoms in the first quadrant. In ideal situations, this nonlinear density response to two successive pulses is quantized, and the quantization value probes the Euler characteristic of the local Fermi sea at the trap center. We investigate the practical effects in experiments, including finite pulse duration, finite edge width of pulses, and finite temperature, which can lead to deviation from quantization. We propose a method to reduce the deviation by averaging measurements performed at the first and third quadrants, inspired by symmetry considerations. With this method, the quantized nonlinear response can be observed reasonably well with experimental conditions readily achieved with ultracold atoms. 
\end{abstract}

\maketitle

Topological physics lies at the center of modern condensed matter physics \cite{Hasan10,Qi11,Bernevig,Witten16,Wen17,Armitage18,Moessner21}. Transport is the most direct tool to detect topological numbers. So far, quantized conductance is mainly limited to one-dimensional metallic systems, and the conductance is quantized to the number of tunneling channels \cite{Landauer57,Fisher81}. This quantized conductance has been observed in both condensed matter systems \cite{vanWees88,Wharam88,Honda95,Frank98,vanWeperen13} and ultracold atomic gases \cite{Krinner15,Krinner17,Lebrat19}.  A two-dimensional band insulator with nontrivial topology hosts stable metallic edge states, whose quantized conductance reveals the topological invariant of the insulating bulk \cite{Hasan10,Qi11,Bernevig}. These include the quantum Hall effect \cite{vonKlitzing80,Thouless82}, the quantum anomalous Hall effect \cite{Haldane88,Chang13}, and the quantum spin Hall effect \cite{Kane05,Bernevig06,Konig07}.  

In a recent paper, Kane proposed a novel effect, which measures a topological invariant of the Fermi sea through transport \cite{Kane22,Physics}. 
The novelty of this effect comes from the following aspects. First, this effect concerns a two-dimensional metal instead of a one-dimensional one. Secondly, this effect considers the nonlinear conductance instead of the linear conductance. Thirdly, this effect probes the Euler characteristic $\chi$ of the Fermi sea, a topological invariant that has not been extensively discussed in dimension higher than one before. 

\begin{figure}[t]
    \centering
    \includegraphics[width=0.48\textwidth]{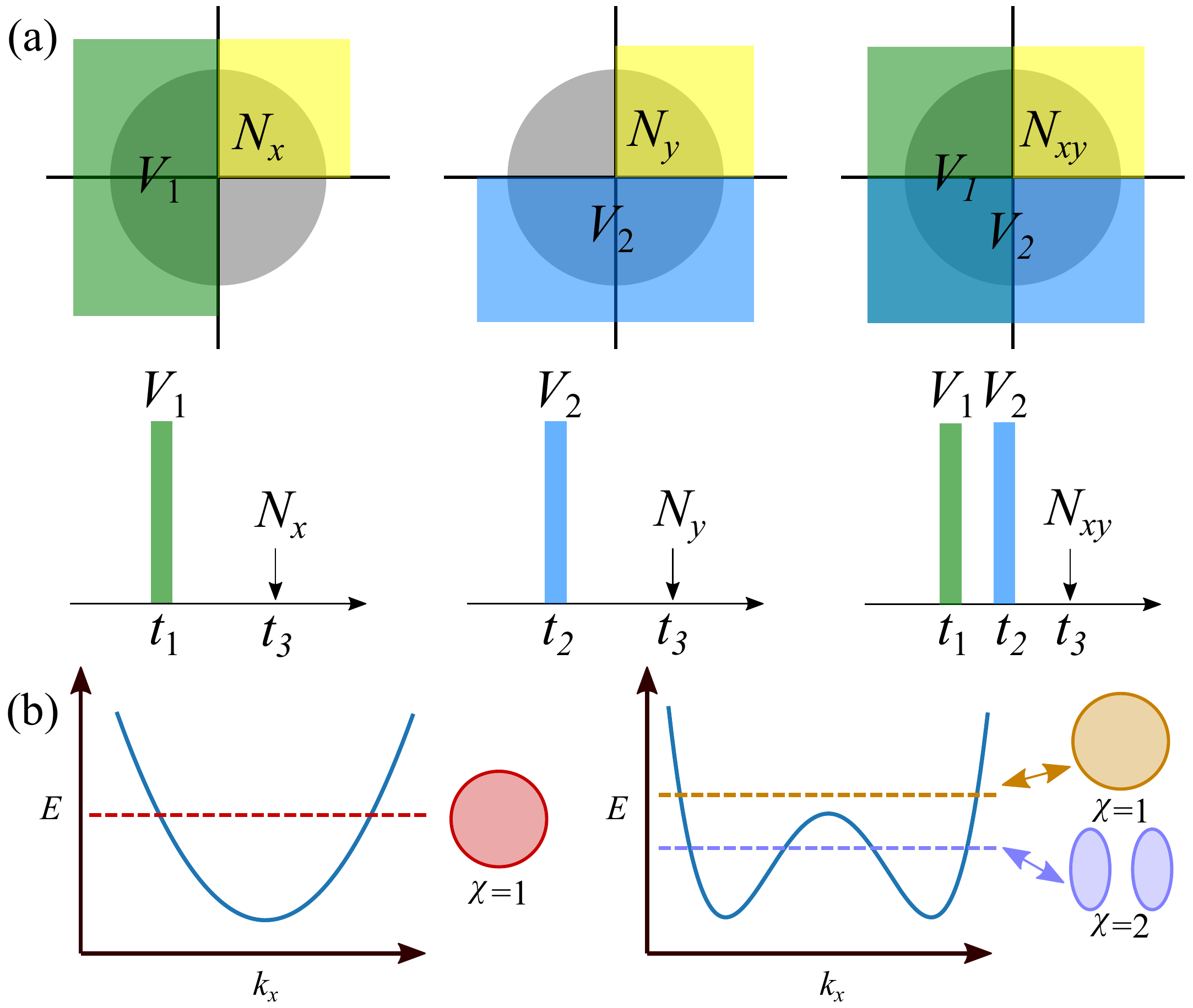}
    \caption{(a) Schematic of the experimental scheme. The green area (left half-plane) and the blue area (lower half-plane) are the areas where two optical pulses are applied. The yellow area (the first quadrant) is the regime where the number of excess atoms is measured. We set $t_1<t_2<t_3$. (b) Schematic of various dispersions and Fermi seas with different Euler characteristics. The circles/ellipses illustrate the topology of the Fermi sea.}
     \label{scheme}
\end{figure}

Formally, the Euler characteristic $\chi$ is defined as the alternating sum of the Betti numbers \cite{Schwartz}
\begin{equation}
\chi=\sum_{n=0}^d(-1)^nb_n,
\end{equation}
where the $n$th Betti number $b_n$ is the rank of the $n$th homology group, and $d$ is the dimension of the manifold. In Morse theory, the Euler characteristic can be computed by evaluating the Hessian at each critical point. In the case of the Fermi sea, we consider a dispersion $E({\bf k})$ filled up to the Fermi energy, $\chi$ can be expressed more concretely using Morse theory as \cite{Kane22,Schwartz} (This expression will be used in the Appendix.)
\begin{equation}
\chi=\sum \text{sgn} \det\Bigg[\frac{\partial^2 E}{\partial k_i\partial k_j}\Bigg]\Bigg\lvert_{{\bf v}=0}.
\end{equation}
Here $\text{sgn}$ is the sign function, $\text{det}[\cdot]$ denotes the determinant, and ${\bf v}=\nabla_{\bf k} E/\hbar$. ${\bf v}=0$ denotes critical points (i.e.  local minima, maxima, or saddle points of the dispersion). The summation runs over all these critical points below the Fermi surface. 
In one dimension, $\chi$ counts the number of disconnected components of the Fermi sea.
In two dimensions, Each electron-like, hole-like, and open Fermi surface contributes $+1$, $-1$, and 0 to $\chi$, respectively.
For instance, as shown in Fig. \ref{scheme}(b), $\chi$ always equals one for a simple quadratic dispersion. Adding spin-orbit coupling along one spatial direction changes the dispersion, resulting in $\chi=2$ at low density. As density increases, the system undergoes a Lifshitz transition to $\chi=1$ \cite{Lifshitz60,Wang12}. Another nontrivial $\chi$ realized in ultracold atom system is the Dirac dispersion in the honeycomb lattice \cite{Tarruell12}. When the chemical potential is above the Dirac point, there are two electron-like Fermi pockets, giving rise to $\chi=2$, and when the chemical potential is below the Dirac point, there are two hole-like Fermi pockets, giving rise to $\chi=-2$. In one-dimension, the famous Landauer formula states that the linear conductance of the ballistic transport is quantized to the Euler characteristic. However, for dimensions higher than one, only a recent work \cite{Kane22} discovered that the nonlinear conductance of the ballistic transport is quantized to the Euler characteristic.  

Two methods have been proposed to detect the quantized nonlinear conductance in Ref. \cite{Kane22}. However, neither has been implemented so far. The first method is based on measuring frequency-dependent ac transport, which is, in principle, doable in condensed matter systems. However, it is still an open question whether this effect is robust enough against local disorders and electron interactions that are inevitably present in solids. The second method is a thought experiment, meaning it is nearly impossible to implement in solid-state materials. Here we point out that the ultracold Fermi gases offer an ideal platform to realize this thought experiment. Moreover, since ultracold atom systems can be disorder-free, and the $s$-wave scattering length between atoms can also be tuned to zero to minimize the interaction effect, this fulfills the requirement of ballistic transport. However, the critical issue is whether this effect is robust enough against practical effects in ultracold atom systems, such as the external confinement potential, the temporal and spatial profiles of the optical pulses, and the thermal effect. Addressing these effects becomes the key for ultracold atom observation of this quantization. After considering all these practical effects, we conclude that this quantization can be readily observed within the current capability of ultracold atom experiments. 

\textit{Experimental Scheme.} Before entering the detailed derivations, we first describe our measurement scheme as follows. We start with a non-interacting spinless degenerate Fermi gas in a two-dimensional harmonic trap 
\begin{equation}
V({\bf r})=\frac{1}{2}m(\omega^2_x x^2+\omega^2_y y^2),
\end{equation} 
where $m$ is the mass of atoms, and $\omega_{x}$ and $\omega_y$ are harmonic trap frequencies along $\hat{x}$- and $\hat{y}$-directions, respectively. 

This proposal contains applying two pulses, one applied on the left-half plane with $x<0$ at time $t_1$ and the other applied on the lower-half plane with $y<0$ at time $t_2$, as shown in Fig. \ref{scheme}. Both pulses apply a uniform potential, which read
\begin{align}
V_1&=\xi h \delta(t-t_1)\Theta(-x), \\
V_2&=\xi h \delta(t-t_2)\Theta(-y).
\end{align} 
Here $\Theta(\cdot)$ is the Heaviside step function, and $h$ is the Planck constant. $\xi$ is an integer. For typical ultracold atom experiments, one can choose $\xi$ from several to tens. Ideally, these pulses are $\delta$ functions in the temporal domain and step functions across $x=0$ or $y=0$. In practice, the pulse should have a finite duration time $\sigma_{t}$, and we require that the temporal integration of the potential strength equals $\xi h$. The step function is also smoothed into a domain with edge width $\sigma_{r}$. Here we first present the general scheme and we will investigate the effect of finite $\sigma_{t}$ and $\sigma_{r}$ later.  

\textit{Step One.} Starting from the equilibrium state, we apply a pulse on the left-half plane with $x<0$ at time $t_1$, and then measure the atom number in the first quadrant at time $t_3$, subtracted by $\overline N=N/4$ with $N$ being the total number of atoms,
\begin{equation}
\int_0^\infty dx\int_0^\infty dy n(x,y,t_3)-\overline{N}. \label{number}
\end{equation}
The result denoted by $N_x$. 

\textit{Step Two.} Starting again from the equilibrium state, we apply a pulse on the lower-half plane with $y<0$ at time $t_2$, and then measure the atom number in the first quadrant at time $t_3$, subtracted by $\overline N$ using the same expression as Eq. \ref{number}. We denote this result as $N_y$.

\textit{Step Three.} Starting again from the equilibrium state, we first apply a pulse on the left-half plane with $x<0$ at time $t_1$, and then apply another pulse on the lower-half plane with $y<0$ at time $t_2$. We measure the atom number in the first quadrant at time $t_3$, subtract by $\overline N$ to obtain $N_{xy}$ using Eq. \ref{number}.

Then, we can obtain a value 
\begin{equation}
\chi=\frac{N_{xy}-N_x-N_y}{\xi^2}. \label{chi}
\end{equation} 
Ideally, $\chi$ should be quantized and the quantization value equals the Euler characteristic of the Fermi sea. The practical effects, such as finite temperature, finite $\sigma_{r}$, and finite $\sigma_{t}$, lead to derivation from the quantized value. To reduce this derivation, we can proceed to \text{Step Four}. 

\textit{Step Four.} Repeating \textit{Step One} to \textit{Step Three} with the same pulses, the difference is to measure $N_x$, $N_y$, and $N_{xy}$ in the third quadrant instead of the first quadrant. This yields another $\chi^\prime$ by using Eq. \ref{chi}. Then, we average $\chi$ and $\chi^\prime$, and this leads to a value closer to quantized Euler characteristic.

We note that, according to Eq. \ref{chi}, although $\chi$ should be quantized to a value of the order of unity, the error bar of measured atom numbers $N_x$, $N_y$ and $N_{xy}$ do not need to be of the order of unity. Instead, the error bar only needs to be smaller than $\xi^2$. At the end of this work, we will show that $\xi\sim 5$ is good enough to reach good quantization. Such a high accuracy of measuring atom number is challenging, but it is already within the capability of current ultracold atomic experiments \cite{detection}.

\textit{Proof of the Scheme.} We first prove the scheme for ideal situations. We consider the collisionless Boltzmann equation of noninteracting fermions in a harmonic trap, 
\begin{equation}
\partial_t f+{\bf v}_{\bf k}\cdot\nabla_{\bf r} f+\frac1{\hbar}({\bf F}-\nabla_{\bf r}V)\cdot\nabla_{\bf k}f=0, \label{BoltzmannE}
\end{equation}
where $f({\bf r},{\bf k})$ is the distribution function. 
${\bf v}_{\bf k}$ is the velocity given by $(1/\hbar)\nabla_{\bf k}\epsilon_{\bf k}$. For the purpose of illustration, we first consider the simple situation of the quadratic dispersion $\epsilon_{\bf k}=\hbar^2{\bf k}^2/(2m)$. ${\bf F}$ is the force given by these two pulses, and ${\bf F}={\bf F}_1$ or ${\bf F}_2$ for the first two steps, and ${\bf F}={\bf F}_1+{\bf F}_2$ for the third step. ${\bf F}_1=\xi h \delta(t-t_1)\delta(x)\hat{{\bf x}}$, and ${\bf F}_2=\xi h \delta(t-t_2)\delta(y)\hat{{\bf y}}$. At $t=0$, $f({\bf r},{\bf k})$ is the semi-classical Fermi distribution
\begin{equation}
f_0({\bf r},{\bf k})=n_\text{F}(\epsilon_{{\bf k}}+V({\bf r})-\mu), \label{f0}
\end{equation} 
where $n_\text{F}(E)=1/(e^{\beta E}+1)$ is the Fermi distribution function with $\beta=1/(k_\text{B}T)$ being the inverse temperature. 
It is easy to show that Eq. \ref{f0} satisfies Eq. \ref{BoltzmannE} when ${\bf F}=0$. 

To solve Eq. \ref{BoltzmannE}, we first introduce a mapping between $({\bf r}^\prime,{\bf k}^\prime)$ and $({\bf r},{\bf k})$ as
\begin{align}
&{\bf r}'_i={\bf r}_i \cos(\omega_i \delta t)-\frac{\hbar {\bf k}_i}{m\omega_i}\sin(\omega_i \delta t); \label{rrprime}\\
&{\bf k}'_i={\bf k}_i \cos(\omega_i \delta t)+\frac{m\omega_i {\bf r}_i}{\hbar}\sin(\omega_i \delta t),\label{kkprime}
\end{align}
where $i=x,y$ and $\delta t=t-t'$. Below we will repeatedly use this mapping with different choices of $\delta t$. We will utilize the fact that when ${\bf F}=0$, $f({\bf r}, {\bf k},t)=f({\bf r}^\prime, {\bf k}^\prime, t^\prime)$. This is ensured by the Liouville's theorem, and the energy conservation $\epsilon_{{\bf k}}+V({\bf r})=\epsilon_{{\bf k}^\prime}+V({\bf r}^\prime)$. Hence, when $t<t_1$, $f({\bf r}, {\bf k})=f_0({\bf r}^\prime,{\bf k}^\prime)=f_0({\bf r},{\bf k})$, and the distribution function does not evolve. 

\begin{figure}[t]
    \centering
    \includegraphics[width=0.45\textwidth]{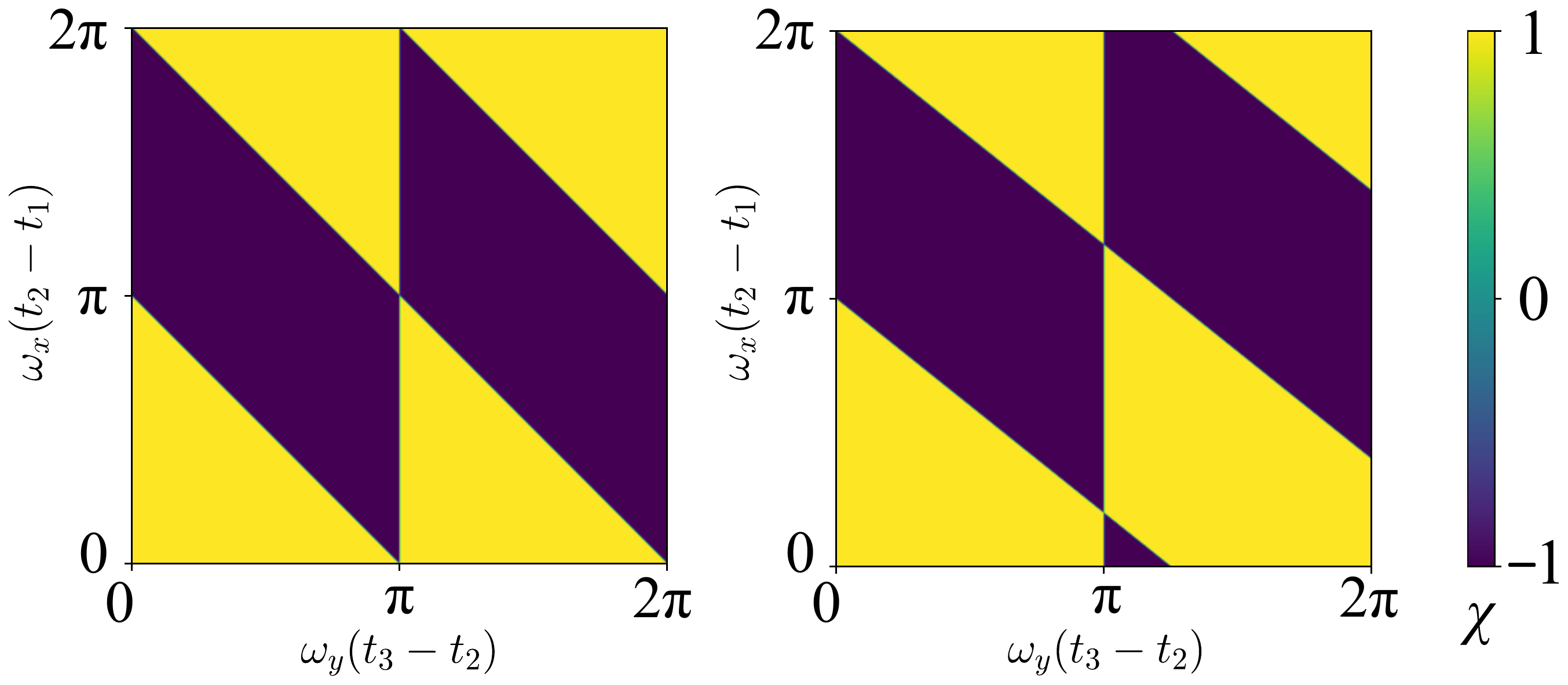}
    \caption{$\chi$ computed from Eq. \ref{chi} following the first three steps for the ideal situation, i.e. pulses with temporal profile of $\delta$-functions and spatial profile of $\Theta$-functions applied to Fermi gas in a harmonic trap at zero temperature. We choose $\omega_x=\omega_y$ for (a) and $\omega_x=0.8\omega_y$ for (b). $t_1$, $t_2$ and $t_3$ are labelled in Fig. \ref{scheme}. }
     \label{ideal}
\end{figure}

For \textit{Step One}, note that the force ${\bf F}_1$ only acts on atoms at the boundary $x=0$ and at an instantaneous time $t_1$. The distribution function changes from $f_0({\bf r}_1,{\bf k}_1)$ to $f_0({\bf r}_1,{\bf k}_1)+ f_1({\bf r}_1,{\bf k}_1)$ at $t=t_1$, and $f_1({\bf r}_1,{\bf k}_1)$ is given by 
\begin{equation}
f_1({\bf r}_1,{\bf k}_1)=-2\pi\xi \delta(x_1)\frac{\partial f_0({\bf r}_1,{\bf k}_1)}{\partial k_{1x}}. \label{f1}
\end{equation}
After that, when we measure $N_x$ at time $t_3$, the distribution function $f({\bf r},{\bf k})$ at $t=t_3$ is given by $f_0({\bf r}_1,{\bf k}_1)+ f_1({\bf r}_1,{\bf k}_1)$, where $({\bf r}_1,{\bf k}_1)$ and $({\bf r},{\bf k})$ are related by the mapping Eq. \ref{rrprime}-\ref{kkprime} with $\delta t=t_{31}=t_3-t_1$. Then, it is easy to see that
\begin{equation}
N_x=\frac{1}{(2\pi)^2}\int_{\text{1q}}d^2{\bf r}\int d^2{\bf k} f_1({\bf r}_1, {\bf k}_1), \label{Nx}
\end{equation}
where ``1q" stands for the first quadrant. Similarly, for \textit{Step Two}, we can obtain that at time $t=t_2$, the distribution function is changed to $f_0({\bf r}_2,{\bf k}_2)+ f_2({\bf r}_2,{\bf k}_2)$, with 
\begin{equation}
f_2({\bf r}_2,{\bf k}_2)=-2\pi\xi \delta(y_2)\frac{\partial f_0({\bf r}_2,{\bf k}_2)}{\partial k_{2y}}.  \label{f2}
\end{equation} 
And we have
\begin{equation}
N_y=\frac{1}{(2\pi)^2}\int_{\text{1q}}d^2{\bf r}\int d^2{\bf k} f_2({\bf r}_2, {\bf k}_2), \label{Ny}
\end{equation}
where  $({\bf r}_2,{\bf k}_2)$ is related to  $({\bf r},{\bf k})$ by Eq. \ref{rrprime}-\ref{kkprime} with $\delta t=t_{32}=t_3-t_2$.

Now we consider \textit{Step Three}. Following the same discussion, it can be shown that $f({\bf r}, {\bf k})$ at time $t=t_3$ contains two parts. The first part is $f_0({\bf r}_1, {\bf k}_1)+f_1({\bf r}_1,{\bf k}_1)$, where $f_1({\bf r}_1,{\bf k}_1)$ given by Eq. \ref{f1} is due to the force ${\bf F}_1$, and $({\bf r}_1,{\bf k}_1)$ and $({\bf r},{\bf k})$ are related by the mapping Eq. \ref{rrprime}-\ref{kkprime} with $\delta t=t_{31}$. The second part is $f_2({\bf r}_2,{\bf k}_2)+f_{12}({\bf r}_2,{\bf k}_2)$ due to the force ${\bf F}_2$, where 
\begin{equation}
f_{12}=(2\pi)^2\xi^2\delta(y_2)\frac{\partial}{\partial k_{2y}}\left[\delta(x_1)\frac{\partial f_0}{\partial k_{1x}}\right],
\end{equation}
where $({\bf r}_2,{\bf k}_2)$ and $({\bf r},{\bf k})$ are related by the mapping Eq. \ref{rrprime}-\ref{kkprime} with $\delta t=t_{32}$, and $({\bf r}_2,{\bf k}_2)$ and $({\bf r}_1,{\bf k}_1)$ are related by the same mapping with $\delta t=t_{21}=t_2-t_1$. Compared with Eq. \ref{Nx} and \ref{Ny}, it is easy to see that $\chi$ defined by Eq. \ref{chi} is given by 
\begin{equation}
\chi=\int_{\text{1q}}d^2{\bf r}\int d^2{\bf k} \delta(y_2)\frac{\partial}{\partial k_{2y}}\left[\delta(x_1)\frac{\partial f_0}{\partial k_{1x}}\right]. \label{chi-f0}
\end{equation}
This counts for the number of extra atoms cooperatively kicked by the two pulses to the first quadrant, that is, they are first kicked by the first pulse to the right side, and then kicked by the second pulse to the first quadrant. Therefore, this is attributed to the nonlinear conductance.

For the harmonic trap, Eq. \ref{chi-f0} can be evaluated exactly and the results acquires a simple expression as  
\begin{equation}
\chi=\text{sgn}[\sin(\omega_x t_{31})\sin(\omega_y t_{32})], \label{chi_harmonic}
\end{equation}
where $\text{sgn}[\cdot]$ stands for the sign function. 
We plot a typical result in Fig. \ref{ideal}. The extra atoms kicked by the pulses evolve in time under the force from the harmonic trap. When $t_{31}<\pi/\omega_x$ and $t_{32}<\pi/\omega_y$, these extra atoms reside in the first quadrant, which result in quantized value $\chi=1$, consistent with the Euler characteristic of the Fermi sea for the quadratic dispersion. For longer evolution time, $\chi$ is either $+1$ or $-1$. This is because inside a harmonic trap, atoms oscillate with the same frequency despite their different velocities.

If the dispersion is not the simple quadratic one, such as in the presence of spin-orbit coupling, or if the trap is not harmonic, the simple expression of Eq. \ref{chi_harmonic} no longer holds. To this end, we consider the situation that the time separation between two pulses is short enough such that we can take ${\bf r}_1\approx {\bf r}_2$ and ${\bf k}_1\approx {\bf k}_2$. We also consider that $t_{31}$ and $t_{32}$ are both sufficiently short. That is to say, if we require $x>0$ at time $t_3$ and $x_1=0$ at time $t_1$, it means $v_{1x}>0$. Similarly, if $y>0$ at time $t_3$ and $y_2=0$ at time $t_2$, it means $v_{2y}>0$. Hence, after the spatial integration we can effectively replace $\delta(x_1)$ and $\delta(y_2)$ by $\Theta(v_{1x})$ and $\Theta(v_{2y})$. Furthermore, by ${\bf v}_1 \approx{\bf v}_2$,    Eq. \ref{chi-f0} recovers the same expression as the uniform situation discussed in Ref. \cite{Kane22}. Following the same derivation presented in Ref. \cite{Kane22}, Eq. \ref{chi-f0} is the Euler characteristic of the two-dimensional Fermi sea. (See Appendix for detailed derivation.) Thus, it can yield other integers of $\chi$ for nontrivial dispersions. To be more precise, note that in an external trap, we can introduce the local Fermi sea determined by the local chemical potential $\mu({\bf r})=\mu-V({\bf r})$. Hence, for short time separation $t_{21}$ and $t_{32}$, Eq. \ref{chi-f0} probes the Euler characteristic of the local Fermi sea at ${\bf r}\approx 0$, which is the intersection between the interfaces of the two pulses. Furthermore, it is easy to see that our scheme can be generalized by varying the areas acted by the two pulses, and we can detect the Euler characteristic of local Fermi sea at different locations.  

\begin{figure}[t]
    \centering
    \includegraphics[width=0.48\textwidth]{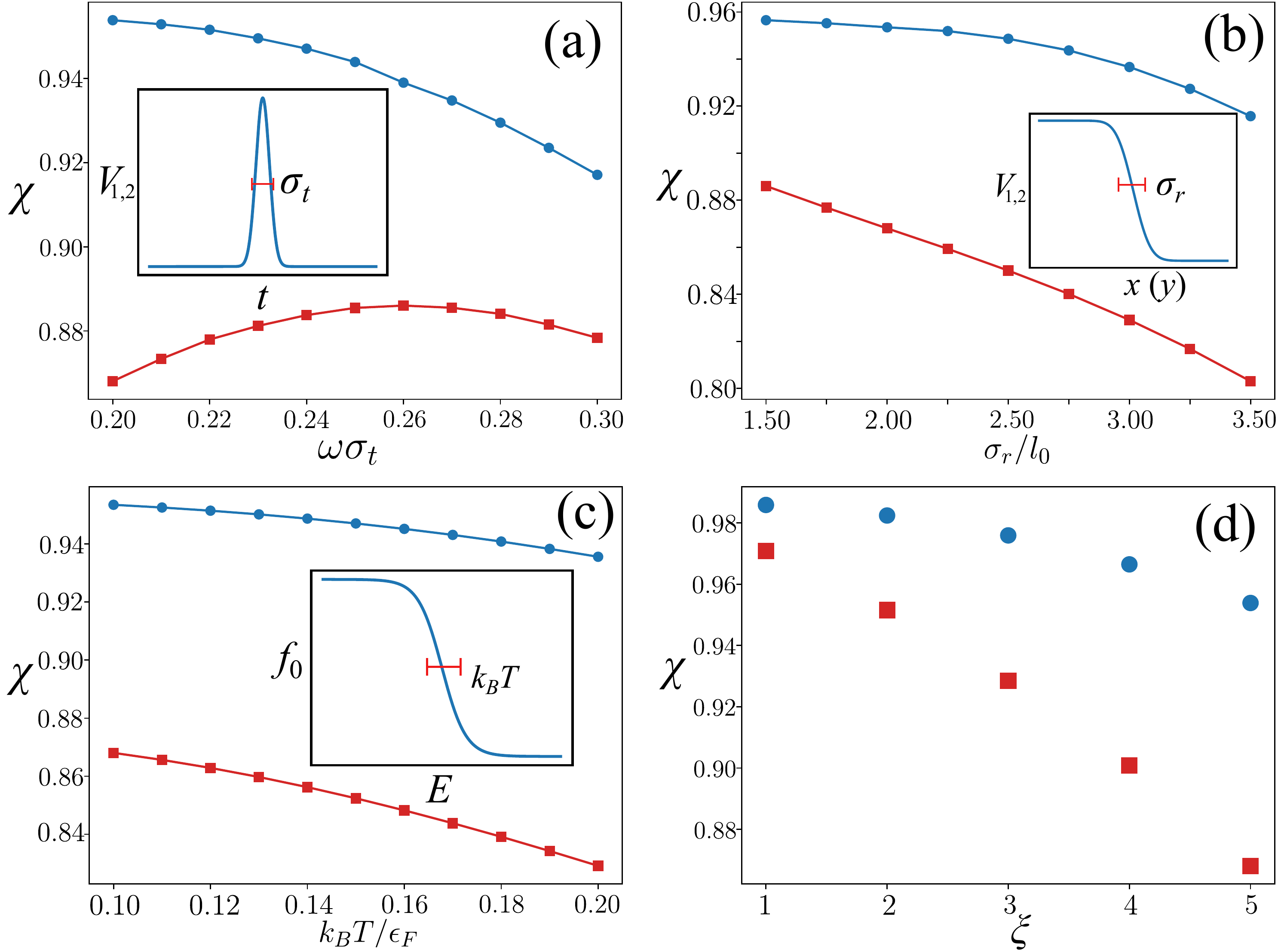}
    \caption{ How $\chi$ is affected by the pulse duration $\sigma_t$, the finite width of the edge $\sigma_r$, and the temperature $T$. Here we fix $\omega_x=\omega_y=\omega$, and $l_0$ is the corresponding harmonic length. (a) $\chi$ as a function of the pulse duration $\sigma_{t}$ (schematically shown in the inset). We fix $\sigma_{r}/l_0=2.0$, $k_\text{B}T=0.1\mu$, and $\xi=5$. (b) $\chi$ as a function of the edge width $\sigma_{r}$ (schematically shown in the inset). We fix $\omega\sigma_{t}=0.2$, $k_\text{B}T=0.1\mu$, and $\xi=5$.  (c) $\chi$ as a function of the temperature $T$. We fix $\omega\sigma_{t}=0.2$ and $\sigma_{r}/l_0=2.0$, and $\xi=5$. (d) $\chi$ for different pulse strength $\xi$. We fix $\omega\sigma_{t}=0.2$, $\sigma_{r}/l_0=2.0$, and $k_\text{B}T=0.1\mu$. The red squares only measure the first quadrant and the blue dots average over the first and the third quadrants.}
     \label{dev}
\end{figure}

By symmetry, if we measure the atom number in the third quadrant, it is equivalent to measuring atom number in the first quadrant but with $\xi$ replaced by $-\xi$ for both pulses. For ideal situations, since $\chi$ is a nonlinear response purely proportional to $\xi^2$, we can conclude that $\chi$ takes the same value for the first and the third quadrant. If we measure the atom number in the second or the fourth quadrants, it equals to the atom number measured in the first quadrant but with $\xi$ replaced by $-\xi$ for one of the two pulses. Thus, $\chi$ takes the opposite value if measured in the second or the fourth quadrants. This is also evident in Fig. \ref{ideal}. Fig. \ref{ideal} shows that $\chi$ is either $+1$ or $-1$. In Fig. \ref{ideal}(a), when $\omega_x=\omega_y$ and when $t_{2}\rightarrow t_{1}$, the dynamical trajectory of atoms always lie in the first and the third quadrants, therefore, $\chi$ is always $+1$.   

\textit{Practical Considerations.} Now we turn to more realistic effects in experiments. We will take into account the effects of finite temperature and finite edge width and duration of pulses. For example, 
we consider a cloud of a total of $N\sim 10^4$ atoms confined to a two-dimensional isotropic trap with frequency $\omega_x=\omega_y=\omega=2\pi\times100$Hz, which corresponds to a harmonic length $l_0=\sqrt{\hbar/(m\omega)}\approx1.6\mu$m for ${}^{40}$K. We set the chemical potential $\mu=100\hbar\omega$, which gives a cloud radius $\approx 14 l_0$. We approximate the $\delta$-functions with Gaussian function, namely
\begin{equation}
\delta(t-t_i)\rightarrow\frac1{\sqrt{2\pi}\sigma_t}e^{-(t-t_i)^2/2\sigma_t^2},
\end{equation}
\begin{align}
&\delta(x)\rightarrow\frac1{\sqrt{2\pi}\sigma_r}e^{-\frac{x^2}{2\sigma_r^2}}, \   \
\delta(y)\rightarrow\frac1{\sqrt{2\pi}\sigma_r}e^{-\frac{y^2}{2\sigma_r^2}}.
\end{align}
We numerically compute $\chi$ defined by Eq. \ref{chi} by solving the collisionless Boltzmann equation \cite{code}. We consider the quadratic dispersion and take $\omega t_1=0.4\pi$, $\omega t_2=0.8\pi$, and $\omega t_3=1.2\pi$. In the ideal situation, this yields $\chi=1$. Considering the practical effects, $\chi$ deviates from unity as shown by red squares in Fig. \ref{dev}. 

In ideal situations, ${\bf F}_{i}$ ($i=1,2$) is a $\delta$-function pulse. Therefore, ${\bf F}_1$ only acts on $f_0$, giving rise to $\ f_1$. ${\bf F}_2$ acts on $f_0$ and $f_1$, giving rise to $f_2$ and $f_{12}$, respectively. In practical situations, because ${\bf F}_i$ has finite duration, ${\bf F}_1$ also influences on $f_1$, and ${\bf F}_2$ also influences on $f_2$ and $f_{12}$. That is to say, $f_{12}$ contains contribution beyond $\xi^2$ order. The high-order corrections is dominated by $\sim\xi^3$ order contribution for small $\sigma_t$ and small $\xi$ and it contributes to a significant part of the total deviation. Due to the symmetry argument presented above, for the $\sim \xi^n$ order contribution with odd $n$, the measurements performed in the first and the third quadrants take opposite values. Hence, we propose a method to reduce the deviation. This method is to measure $\chi$ and $\chi'$ in the first and the third quadrants, respectively, and then average over these two results, as described by {\it Step Four}. The averaged results are shown by blue dots in Fig. \ref{dev}. As one can see, after the average, $\chi$ can be $\sim 0.95$ for very practical experimental conditions, such as $\omega \sigma_{t}=0.2$, $\sigma_{r}/l_0=2.0$, $k_\text{B}T=0.1\mu$, and $\xi=5$. These conditions can be routinely reached within current ultracold atom experimental capability. Moreover, since the applied pulses are a far-detuned weak potential, we expect the atom loss to be negligible in this process. On the other hand, here, we choose the total number of atoms to be $\sim 10^4$, which requires a higher accuracy in measuring atom number. One can also reduce the total atom number, but this will require more stringent requirements on experimental parameters, since a smaller Fermi surface is much more fragile. In a real experiment, one can further optimize these parameters to reach the best balance.

\textit{Summary and Outlook.} Ultracold atomic gas is an ideal platform to experimentally observe the recently proposed quantized nonlinear conductance in Ref. \cite{Kane22}. The key message of this work is to show that this effect can be  observed with practical experimental conditions, calling for immediate experimental implementations. The experimental observation will enable the study of the interaction effect on quantized nonlinear transport in a highly controllable way. The Euler characteristic of Fermi sea is also related to topological multipartite entanglement entropy of Fermi liquids \cite{Tam22}. Further exploration of Euler characteristic can advance our understanding of the role of topology in high-dimensional Fermi liquids and enrich the topological physics.

\textit{Acknowledgment.} We thank Zhe-Yu Shi, Chengshu Li, Yanting Cheng, and Qi Gu for helpful discussions. The project is supported by Beijing Outstanding Young Scholar Program, Innovation Program for Quantum Science and Technology 2021ZD0302005, NSFC Grant No.~11734010 and the XPLORER Prize. F.Y. is also supported by Chinese International Postdoctoral Exchange Fellowship Program (Talent-introduction Program) and Shuimu Tsinghua Scholar Program at Tsinghua University.

\section*{Appendix}

Here we elaborate on the short-time response for an arbitrary dispersion and trap profile. We show that in the short-time limit, one recovers the expression for Euler characteristics derived in Ref. \cite{Kane22}. 

In the limit $t_{31},t_{32}\to 0$, we can linearize the trajectory of motion
\begin{align}
&x_1\approx x-t_{31}v_{1x},\\
&y_2\approx y-t_{32}v_{2y},
\end{align}
and use ${\bf k_1}\approx {\bf k_2}\approx {\bf k}$ in the derivatives.
In this way, we obtain an asymptotic form of $\chi$,
\begin{multline}
\chi=\int_{\text{1q}}d^2{\bf r}\int d^2{\bf k} \delta(y-t_{32}v_{2y})\\
\frac{\partial}{\partial k_{y}}\left[\delta(x-t_{31}v_{1x})\frac{\partial f_0}{\partial k_{x}}\right]. 
\end{multline}
Next, we perform the integration over the 1st quadrant. The $\delta$-functions put a constraint on the velocity and generate two step functions $\Theta(v_{2y})$ and $\Theta(v_{1x})$, giving
\begin{equation}
\chi=\int d^2{\bf k}\Theta(v_{2y})\frac{\partial}{\partial k_y}\Bigg[\Theta(v_{1x})\frac{\partial f_0}{\partial k_x}\Bigg]
\end{equation}
Further utilizing ${\bf v_1}\approx{\bf v_2}\approx{\bf v}$ and performing integration by parts, we arrive at
\begin{equation}\label{lim}
\chi=-\int d^2{\bf k}\Theta(v_x)\frac{\partial\Theta(v_y)}{\partial k_y}\frac{\partial f_0}{\partial k_x}.
\end{equation}
The above equation, which is obtained in the short time limit $t_{31},t_{32}\to0$ for arbitrary dispersion and trap profile, is the same as the expression for the uniform case in Ref. \cite{Kane22}.

In the rest of the Appendix, we follow Ref. \cite{Kane22} and show that Eq. \ref{lim} is indeed the Euler characteristic of the Fermi sea. 
First one can add a null term to Eq. \ref{lim}
\begin{equation}
\int d^2{\bf k}\Theta(v_x)\frac{\partial\Theta(v_y)}{\partial k_x}\frac{\partial f_0}{\partial k_y}.
\end{equation} 
This term can be rewritten using the chain rule as 
\begin{equation}
\begin{split}
&\int d^2{\bf k}\Theta(v_x)\frac{\partial\Theta(v_y)}{\partial v_y}\frac{\partial v_y}{\partial k_x}\frac{\partial f_0}{\partial E}\frac {\partial E}{\partial k_y}\\
&=\int d^2{\bf k}\Theta(v_x)\frac{\partial v_y}{\partial k_x}\frac{\partial f_0}{\partial E}\hbar v_y\delta(v_y).
\end{split}
\end{equation}
The term $v_y\delta(v_y)$ in the integrand ensures the null term to be zero.

After adding the null term and performing integration by parts, one gets
\begin{equation}
\begin{split}
\chi&=\int d^2{\bf k} f_0\Bigg[\frac{\partial\Theta(v_x)}{\partial k_x}\frac{\partial\Theta(v_y)}{\partial k_y}-\frac{\partial\Theta(v_x)}{\partial k_y}\frac{\partial\Theta(v_y)}{\partial k_x}\Bigg]\\
&=\int d^2{\bf k} f_0\delta({\bf v})\Bigg[\frac{\partial v_x}{\partial k_x}\frac{\partial v_y}{\partial k_y}-\frac{\partial v_x}{\partial k_y}\frac{\partial v_y}{\partial k_x}\Bigg].
\end{split}
\end{equation}
Because of the factor $f_0$ and $\delta({\bf v})$, only points below the Fermi surface with ${\bf v}=0$ contribute to the integration.
One can then perform the change of variables by replacing the momentum integration with a velocity integration, which introduces a Jacobian determinant
\begin{equation}
J=\det\Bigg[\frac{\partial(k_x,k_y)}{\partial(v_x,v_y)}\Bigg],
\end{equation}
giving
\begin{equation}
\chi=\int d^2{\bf v} f_0\delta({\bf v})|J|\Bigg[\frac{\partial v_x}{\partial k_x}\frac{\partial v_y}{\partial k_y}-\frac{\partial v_x}{\partial k_y}\frac{\partial v_y}{\partial k_x}\Bigg],
\end{equation}
with $|J|$ the absolute value of $J$. 

Noticing that the terms inside the bracket is simply the inverse of the Jacobian determinant, we get
\begin{equation}
\chi=\sum \text{sgn}\det\Bigg[{\frac{\partial(v_x,v_y)}{\partial(k_x,k_y)}}\Bigg]\Bigg\lvert_{{\bf v}=0},
\end{equation}
where the summation runs over all critical points with ${\bf v}=0$ below the Fermi surface. Finally, utilizing ${\bf v}=\nabla_{\bf k}E/\hbar$, we arrive at the expression
\begin{equation}
\chi=\sum \text{sgn} \det\Bigg[\frac{\partial^2 E}{\partial k_i\partial k_j}\Bigg]\Bigg\lvert_{{\bf v}=0}.
\end{equation}
This is exactly the formula for the Euler characteristics in the Morse theory \cite{Kane22,Schwartz}.

\end{document}